\documentclass[a4paper,11pt]{article}
\usepackage{jheppub} % for details on the use of the package, please
\usepackage{amsmath,amssymb,amsfonts}
\usepackage{mathtools}
\usepackage{amsfonts}
\usepackage{microtype}
\usepackage{setspace}
\usepackage{lineno,hyperref}
\usepackage{array,multirow,graphicx}
\usepackage{float}
\usepackage[font=scriptsize]{caption}
\usepackage[T1]{fontenc} % if needed
\usepackage{color,soul}

\title{Spinning Toroidal Brane Cosmology;\\ A Classical and Quantum Survey}

\author[a,c]{Salman Abarghouei Nejad,}
\author[b,1]{Mehdi Dehghani,\note{Corresponding author.}}
\author[c]{and Majid Monemzadeh}

\affiliation[a]{Departamento de Física, Universidade Federal de Pernambuco, 50670-901, Recife, Braszil}
\affiliation[b]{Department of Physics, Faculty of Science, Shahrekord University, 115, Shahrekord, Iran}
\affiliation[c]{Department of Physics, University of Kashan, 87317-51167, Kashan, Iran}

\emailAdd{salmanabar@df.ufpe.br}
\emailAdd{dehghani@sku.ac.ir}
\emailAdd{monem@kashanu.ac.ir}

\abstract{
We construct a cosmological model based on a free particle model which is constrained on an embedded toroidal brane, with a general rotation around a specific axis in the bulk space. Some related issues such as the rotation axis of the brane, the presence of gravitomagnetic background and its relation to the general angular velocity of the brane, and its quantum mechanics and related issues such as minimal length and minimal momentum of the quantum model in the $ \mathbb{T}^3 $ brane are studied. Also, some cosmological features such as the constraint which is imposed upon the toroidal universe  by the isotropy and homogeneity conditions, the corresponding Hubble law, and accelerating expansion for the spinning toroidal model without considering a cosmological constant are also studied.}

\keywords{Brane Cosmology, Spinning Toroidal Universe, Symplectic Structure, Minimal Length, Hubble Law}

\begin{document} 
\maketitle
\flushbottom

\section*{Introduction}

Global properties of the space is not restricted by general relativity, and a universe with a specific local geometry may have various global topologies \cite{phys-rep}. Cosmological models such as Friedmann-Robertson-Walker (FRW) models allow the universe to have the same average local properties everywhere, while there would be a freedom to describe a quite different space for the large scale. This is due to the fact that by having a negative or vanishing local curvature, one can not necessarily conclude that our universe is infinite, since the universe still can have a finite volume due to the global topological multi-connectivity feature of its space \cite{planck13,planck14}. 

On the other hand, flat or hyperbolic FRW solutions can describe multi-connected universes. Hence, within the inflationary scenarios, quantum fluctuations can produce compact spaces of constant curvature, both from flat spaces \cite{Zeldovich}, and curved ones \cite{Coule,Linde}. This inspires us to think about a multi-connected finite universe, which the simplest one is the torus \cite{Riazuelo,Riazuelo2}.

But, this is not the whole story. The topology of the universe is not the only thing which is missed in standard cosmological models. Particularly, the spinning universe around the specific axis is not physically desirable \cite{Saadeh}, and it would challenge inflation theory that disagrees the  isotropy and homogeneity features of the universe.

On the other hand, some observational evidences such as the temperature anisotropy of the cosmic microwave background (CMB) radiation \cite{Cheng}, the polarization of quasars \cite{Hutsemekers,Hutsemekers2,Hutsemekers3,Hutsemekers4}, the acceleration of the cosmic expansion \cite{Antoniou,cai,Antoniou2,Antoniou3,Antoniou4,Javanmardi}, and angular distribution of the fine-structure constant \cite{Webb,Webb2,Webb3}, may indicate preferred directions in the sky. An important observational evidence obtained by studying the Spiral galaxies with a well-defined handedness in a particular region of the sky shows that the universe may spin on a specific axis \cite{Longo,Longo2}.

In this article, we model a spinning toroidal universe and study its features, while the non-rotating toroidal universe is studied before \cite{torus}. This would be important since a spinning universe has not been seriously studied via a cosmological theory. 

The tool which is used here is the geometrical embedding approach, where  the available configuration space for the test particle is embedded in an extended one, and the corresponding physical and geometrical results can be explained by the formalism of constrained systems.

It is worth mentioning that the most approaches dealing with embedding procedures are concentrated on the tools of Riemannian and pseudo-Riemannian geometries and their corresponding induced metrics. Although this work is based on a similar procedure, instead of inducing  space(time) metric, phase space metric (symplectic tensor) is induced \cite{nakahara}.

Now, by adding the spin to the embedded space in $ \mathbb{R}^4$, we continue the work \cite{torus}, where ADD extra dimensions problem  \cite{nima, ADD}, and making a particle model without referring to general relativity \cite{Maartens,papa}, is concerned. In that article, from the induced metric on a test particle's phase space, a gauge symmetry structure (symplectic tensor with gauge degrees of freedom)  is made, in a way where the extra dimensions of the extended space changed to physical ones (or at least gauged ones).

Here, we follow the different line, i.e. we seek the induced phase space metric on the spinning  $ \mathbb{T}^3 $ in  $ \mathbb{R}^4$ to make a classical and quantum brane cosmology. 

This article is organized in the following way. In the first two sections, we make our spinning toroidal universe and extract its symplectic structure. The third section is dedicated to the features which satisfy the isotropy and homogeneity conditions of the spinning toroidal universe. In section four, we deal with quantum cosmological investigations, and quantum mechanics of the model and some of corresponding features such as minimal length are studied. Last section is  devoted to the cosmic dynamics, where the Hubble law and the expansion acceleration of the model is calculated. Some phenomenological and numerical investigations are also suggested to verify the model.

%most proposed theories of quantum gravity predict topology-change in the early Universe which could be visible at large scales today.

\section{Spinning toroidal universe}

From brane cosmological point of view, this model is built in the form of a $ \mathbb{T}^3 $ brane which is embedded in the Euclidean space $ \mathbb{R}^4 $. Using the theory of constrained systems, we know how to surf the toroidal brane as the boundary of a 3-dimensional universe\footnote{We emphasis that in this article we use the word "boundary" as the 3-dimensional hypersurface of the universe which is embedded in the bulk space.} \cite{torus}. Thus, we only need to add the general spin feature, $ \omega $, to the $ \mathbb{T}^3 $ universe, and consequently to our test particle, living on such a brane. 

In corresponding cosmological coordinates, all points can be described by Cartesian coordinates $ (x,y,z,s) $ relating to the spherical coordinates as,
\begin{eqnarray}
&& x = r \cos \varphi , \nonumber \hspace{2.7 cm} y = r \sin \varphi \cos \psi , \nonumber \\
&& z  = r \sin \varphi \sin \psi \cos \chi , \nonumber \hspace{1cm} s = r \sin \varphi \sin\psi \sin \chi ,
\end{eqnarray}
where, $ 0\leq\phi,\psi <\pi$, $ 0\leq\chi<2\pi $, with $ r \geq 0 $.

 Natural basis vectors of these coordinates are, 
\begin{eqnarray}
&& h_r = 1 , \hspace{1cm} h_\varphi = r ,  \hspace{1cm}  h_\psi = r  \sin\varphi , \hspace{1cm} h_\chi = r \sin\varphi \sin\psi . \nonumber
\end{eqnarray}
Unit vectors are related via $ \hat{e}_i = \frac{\partial \vec{r}}{h_i \partial q_i} $ as,
\begin{eqnarray}
&& \hat{r} = \hat{i} \cos\varphi + \hat{j} \sin\varphi \cos\psi + \hat{k} \sin\varphi \sin\psi \cos\chi + \hat{s} \sin\varphi \sin\psi \sin\chi , \nonumber \\
&& \hat{\varphi} = - \hat{i} \sin\varphi + \hat{j} \cos\varphi \cos\psi + \hat{k} \cos\varphi \sin\psi \cos\chi + \hat{s} \cos\varphi \sin\psi \sin\chi , \nonumber \\
&& \hat{\psi} = - \hat{j} \sin\psi + \hat{k} \cos\psi \cos\chi + \hat{s} \cos\psi\sin\chi , \nonumber \\
&& \hat{\chi} = - \hat{k} \sin\chi + \hat{s} \cos\chi .
\end{eqnarray}
Using the above mentioned unit vectors, one can stablish the velocity vector of the test particle as,
\begin{eqnarray}
 \vec{v} =  \frac{d\vec{r}}{dt}   = \hat{r} \dot{r}+ \hat{\varphi} r \dot{\varphi} + \hat\psi r \dot{\psi} \sin\varphi + \hat{\chi} r \dot{\chi} \sin\varphi \sin\psi .
\end{eqnarray}
Now, we define the velocity vector of the test particle, living on the spinning $ \mathbb{T}^3 $ in $ \mathbb{R}^4 $ as,
\begin{eqnarray}
\vec{v}_{\mathsf{tot}}=\vec{v}+\vec{v}_{\omega}.
\end{eqnarray}
The direction of the $ \vec{v}_{\omega} $ should be chosen in such a way, not to enter a  fundamental or coordinate singularity to the kinetic energy (or the Hamiltonian) function, and at the same time, add the fundamental constant $ \omega $ (or a new Casimir of the corresponding algebra) to the model. For instance, adding the spin velocity term in the form of $ \vec{v}_{\omega} = \hat{\psi} r \sin\varphi \omega  $, which results to the term $ v_{\psi} =r\sin\varphi \dot{\psi} +v_{\omega} $, enters a coordinate singularity in Hamiltonian, which is not desirable. We ignore such an addition to avoid singularity problems. Whereas, one can add the angular velocity (spin) $ \vec{\omega} = \omega \hat{\psi} $ to the test particle's spin, $ (\omega + \dot{\psi}) $, without entering any singularity to the Hamiltonian.

Hence, the test particle's Lagrangian with unit mass living on such a universe is explained as,
\begin{equation}
L = \frac{1}{2}(\dot{r}^2 + r^2 \dot{\varphi}^2 + r^2(\omega + \dot{\psi})^2 \sin^2\varphi + r^2 \dot{\chi}^2 \sin^2\varphi \sin^2\psi),
\end{equation}
and corresponding velocities are,
\begin{eqnarray}
&& \dot{r} = p_r, \hspace{1cm} \dot{\varphi} = \frac{p_\varphi}{r^2}, \hspace{1cm} \dot{\psi} = \frac{p_\psi}{r^2 \sin^2\varphi}-\omega,  \hspace{1cm} \dot{\chi} = \frac{p_\chi}{r^2 \sin^2\varphi \sin^2\psi}.
\end{eqnarray}
Having Legendre Transformation, the energy function of the test particle on the embedded brane which is spinning around $ \hat{\psi} $, can be obtained as,
\begin{eqnarray}\label{Hamiltonian}
 H_c  = \frac{1}{2} (p_r^2 + \frac{p_\varphi^2}{r^2} + \frac{p_\psi^2}{r^2 \sin^2\varphi}+ \frac{p_\chi^2}{r^2 \sin^2\varphi \sin^2\psi}) - \omega p_\psi.
\end{eqnarray}
Here, the term in parenthesis includes a removable  regular coordinate singularity, although the added term would be remained secure.

By redefining corresponding momenta, the Hamiltonian \eqref{Hamiltonian} can be redefined as the Hamiltonian of a free massive particle which is minimally coupled to a gauge potential as,
\begin{equation}
H_c = \frac{1}{2}(\vec{P} - \vec{G})^2 + V ,
\end{equation}
where, $  \vec{G} = \hat{\psi}\omega r \sin\varphi $, and $ V(r,\varphi) = -\frac{1}{2}r^2 \omega^2 \sin^2\varphi $, are 1-form  and  scalar gravitomagnetic potential respectively. The negative nature of this potential, which is similar to the confinement potentials for quarks in QCD, confines the test particle on $ \mathbb{T}^3 $.

The direct tensor product of this vector potential and a suitable differential operator,  $ d_{I} $, one can obtain the background magnetic 2-form of the model as,
\begin{eqnarray}
 B_{IJ} = \frac{1}{2} d_{I} G_{J}, \nonumber
\end{eqnarray}
where,
\begin{eqnarray}
d_{I} := (\partial_r,  \frac{1}{r} \partial_\varphi ,  \frac{1}{r \sin\varphi} \partial_\psi ,  \frac{1}{r \sin\varphi \sin\psi} \partial_\chi) .
\end{eqnarray}
Therefore,
\begin{eqnarray}
B_{IJ} =  \omega ( \hat{r} \otimes \hat{\psi} \sin\varphi + \hat{\varphi} \otimes \hat{\psi} \cos\varphi ),
\end{eqnarray}
will not have any coordinate singularity. Putting the gravitomagnetic pseudovector components in an array, 
\begin{eqnarray}
B_{IJ}:= \omega \begin{pmatrix}
 0 & 0 & \sin\varphi & 0 \\
0 & 0 & \cos\varphi & 0 \\
-\sin\varphi & -\cos\varphi & 0 & 0\\
 0 & 0 & 0 & 0 \\
\end{pmatrix},
\end{eqnarray}
indicates that the presence of $ \omega $ produces a scalar and a magnetic vector potential, which  can be interpreted as the magnetic background resulting the constant noncommutativity, which is consistent to the noncommutativity, obtained from the states of string theory \cite{SW}.
 This magnetic background changes the symplectic structure of the phase space of this model to the  the symplectic structure of a noncommutative one, where a part of this noncommutativity is controlled via $ \omega $, and the other part, as it has been shown in \cite{torus}, is due to the Casimirs of the model which are two radii of  $ \mathbb{T}^3 $, embedded in the Euclidean space $ \mathbb{R}^4 $. 

Also, as we see that the components of the magnetic 2-form depends on a cosmological coordinate, $ \varphi $, with the spatial directions as, $ \hat{r} \otimes \hat{\psi}  $ and $ \hat{\varphi} \otimes \hat{\psi} $.  

 If we neglect the direction along $ \psi $ by imposing the constraints, $ \vec{B} $ can be written in a suitable form as,
 \begin{eqnarray}
 \vec{B} = \omega \sin\varphi \hat{r} + \omega \cos\varphi \hat{\varphi},
 \end{eqnarray}
with the norm of, $ \parallel \vec{B} \parallel = \omega $, which is the product of fundamental spin and unit mass, as the gravitational charge current. Thus, one can suggest that due to the spin of the universe, from the $ \mathbb{R}^4 $ observer's point of view, for every massive particle, $ m $, there is a gravitomagnetic monopole with the vector charge $ g = m \vec{\omega} $ \cite{mash}. From $ \mathbb{T}^3 $ observer's point of view\footnote{The visible spin direction, which can be measured by $ \mathbb{T}^3 $ observer is defined as  the vector $  \vec{\Omega}_{\mathsf{vis}}^{(\mathbb{T}^3)}  $. For details the reader should refer to the section \ref{Toward}. These monopoles and their interactions are also studied in \cite{jadid}.}, this gravitomagnetic monopole can be measured as $ \vec{g}_{\mathsf{vis}}^{(m)} = m  \vec{\Omega}_{\mathsf{vis}}^{(\mathbb{T}^3)}  $.

\section{Symplectic structure of spinning toroidal universe}
By studying the symplectic structure, the dynamics of the particle's motion, the geometry of the phase space, symmetry generators' Lie algebra of the test particle's motion, and consequently symmetry generators' algebra of phase and configuration spaces are accessible.

In the $ \mathbb{R}^4 $ observer's cosmological coordinates, the brane of our toroidal universe is described with the following constraint,
\begin{equation}\label{phi1bar}
\Phi_1=r^2 - \varsigma_+^2 - \varsigma_-^2 - 2 \varsigma_- \varsigma_+\cos\psi,
\end{equation}
where, $ r^2=x^2+y^2+z^2+s^2 $, and $ \varsigma_{+}$, and $ \varsigma_{-}$ are 3-torus' radii. 

The constraint chain structure of the model includes one more constraint. In other words, the effect of the spin in the phase space appears due to the change in the Hamiltonian as,
\begin{equation}
\Phi_2 = \{ \Phi_1, H\}= 2 r p_r + \frac{2  \varsigma_- \varsigma_+ \sin\psi}{r^2 \sin^2\varphi}(p_\psi - r^2 \omega \sin^2\varphi).
\end{equation}
Here, Poisson bracket matrix of constraints, which is the fundamental part of the symplectic structure of the model, does not include $ \omega $.
\begin{equation}
\Delta_{12} = 4 r^2 + (\frac{2  \varsigma_- \varsigma_+ \sin\psi}{r \sin\varphi})^2.
\end{equation}
 Therefore, in comparison to the non-spinning toroidal universe \cite{torus}, the spin parameter of the universe appears linearly in the symplectic structure.
 
If we can be assured of two basic cosmological principles, i.e. the isotropy and homogeneity features of the universe, one can introduce the cosmic $ \mathbb{R}^4 $  observer as the embedded $ \mathbb{T}^3 $ observer either. This can be done by calculating Dirac brackets and using coordinate reduction methods.

One should notice that if two radii of our toroidal universe is considered to be nearly equal, then our model can satisfy the homogeneity condition by approximation. Also, if we can find the non-unique spin direction, including some free parameters, we can think that our model can be estimated as an isotropic universe.

As we know, by considering two constraints of the phase and configuration spaces, Dirac brackets of basic components of phase space  as $ z_\mu := (r,\varphi,\psi,\chi)\oplus(p_r,p_\varphi,p_\psi,p_\chi) $ can be obtained as,
\begin{eqnarray}\label{dirac_b}
\{z_\mu ,z_\nu\}^* = \{z_\mu ,z_\nu\} + \frac{1}{\Delta_{12}}(\{z_\mu,\Phi_1\}\{\Phi_2,z_\nu\}-\mu\leftrightarrow \nu).
\end{eqnarray}
Now, by imposing the constraints and calculating Dirac brackets, one can redefine the following couple of variables,
\begin{eqnarray}
&& \psi = \cos^{-1}(\frac{r^2 - \varsigma_+^2 - \varsigma_-^2}{2 \varsigma_- \varsigma_+}), \nonumber \\
&& p_\psi = r^2 (\omega - \frac{2 r p_r}{((\varsigma_-+\varsigma_+)^2-r^2)(r^2-(\varsigma_+- \varsigma_-)^2)})\sin^2\varphi ,
\end{eqnarray}
where can be ignored to make a 3-dimensional coordinate, i.e. $ r,\varphi $ and $ \chi $, for our  3-dimensional test particle. This procedure  imports the constraints into the dynamics of the system consistently, i.e. the reduced physical phase space is not separated from the $ \mathbb{R}^4 \oplus \mathbb{R}^4 $  initial phase space. The initial phase space is important, since it determines the limited domain available for the remained parameters of the brane, such as $ r $,
\begin{eqnarray}
\varsigma_+ - \varsigma_- \leq r \leq \varsigma_+ + \varsigma_-,
\end{eqnarray}
and also it determines the metric (and the symplectic structure) of the smaller phase space.  For instance, in initial phase space $ \mathbb{R}^4 \oplus \mathbb{R}^4 $, Poisson brackets matrix of basic parameters is obtained as the following symplectic matrix,
\begin{eqnarray}
\{z_\mu,z_\nu\} = J_{\mu\nu}(z)\qquad , \qquad J_{\mu\nu}(z) = \left(
  \begin{array}{cc}
    0 & \textbf{I} \\
    -\textbf{I} & 0 \\
  \end{array}
\right),
\end{eqnarray}
which indicates the canonical structure of phase and configuration spaces. But, after reducing the phase space to $ \mathbb{T}^3 \oplus \mathcal{M} $, and inducing the symplectic structure (phase space metric), we obtain a new $ J_{\mu\nu}(z) $ which depends on the phase space variables. Due to special form of the constraints, $ \Phi_1 $ and $ \Phi_2 $, and their Poisson brackets matrix, we will see that the geometry of the phase space is noncommutative and non-canonical, although the geometry of the configuration space remains commutative. 

Using  \eqref{dirac_b}, we derive non-vanishing Dirac brackets as,
\begin{eqnarray}\label{JS}
&& \{  {r}, {p}_r \}^*  =  \left( 1 -  {r}^2 (1+\epsilon) \sin^2 \varphi \right) ,\nonumber \\
&& \{  {r},p_\psi \}^*  =  \frac{- i \hslash}{2}  (2+\epsilon)   {r}^2 \sin^2 \varphi , \nonumber \\
&& \{ \psi, {p}_r\}^* =-  \left(1 + \frac{1}{2} \epsilon + \frac{1}{8}  {r}^2 (1+\frac{3}{2}\epsilon) (1 - 8 \sin^2 \varphi)  \right), \nonumber \\
&& \{\psi,p_\psi\}^* =  (1 + \epsilon)  {r}^2 \sin^2\varphi , \nonumber \\
&& \{ {p}_r,p_\psi\}^* =   \left( (1+\frac{1}{2}\epsilon) ({r} {p}_r+ {p}_r {r}) - 2 {r} \varsigma^2 \omega \right) \sin^2 \varphi + O({r}^3 {p}_r),   \nonumber \\ 
&& \{ {p}_r,p_\varphi\}^* =   {r}^2  (1 + \frac{1}{2} \epsilon )\omega \sin 2\varphi , \nonumber \\
&& \{p_\psi,p_\varphi\}=    {r}^2 \varsigma^2 \omega \sin 2\varphi + O( {r}^2  {p}_r) .
\end{eqnarray}

Using the canonical Hamiltonian and the obtained symplectic structure, one can calculate every dynamical variable such as ($ r $, $ \dot{r} $), and then move on the constrained surface ($ \mathbb{T}^3 $ hypersurface in configuration space and $ \mathbb{R}^3 $ hypersurface in phase space).

\section{Isotropy and Homogeneity conditions }
\subsection{Isotropic spinning torus universe}

In previous part, we found the spin direction of the 3-dimensional universe. Since this direction is located in the coordinates of a 4-dimensional universe, a 3-dimensional observer living on the toroidal universe cannot perceive this direction completely. Thus, in order to find the portion of the direction of the spin, which is seen by the 3-dimensional observer, we follow the following root.

As it is mentioned, after obtaining the dynamical equations and algebraic calculations, one can move on the constrained surface of $ \Phi_1 $ and  $ \Phi_2 $, and omit the couple variable $ (\psi,p_\psi) $. But, one should notice that such omitted variables cannot be considered as completely nonphysical ones, since the invisible spin of the universe is directed toward $ \psi $. Hence, this direction and its corresponding variable should not be ignored completely.

On the other hand, if we want to ignore a canonical couple for the sake of the second order constraints $ \Phi_1 $ and $ \Phi_2 $, one should have them symplected in the formalism of constrained systems. Hence, to sacrifice a canonical couple to reduce the configuration space from $ \mathbb{R}^4 $ to $ \mathbb{T}^3 $, we redefine the hypersurface $ \mathbb{T}^3 $, embedded in $ \mathbb{R}^4 $ as,
\begin{eqnarray}
\Phi_1(r,\psi) = \psi - \cos^{-1}(\frac{r^2-\varsigma_{+}^{2}-\varsigma_{-}^{2}}{2\varsigma_{+}\varsigma_{-}}).
\end{eqnarray}
 Hence, the visible direction of such a universe tangents to this hypersurface and the direction in the configuration space (the coordinate part of the phase space), which is invisible for the $ \mathbb{T}^3 $ observer, is perpendicular to this hypersurface as $ \hat{n}_{\mathsf{invis}} = \frac{\nabla \Phi_1}{ \mid \nabla \Phi_1\mid } $,
\begin{equation}\label{invis}
\hat{n}_{\mathsf{invis}} = \frac{r^2 \sin\varphi \hat{r} + \varsigma_+ \varsigma_- \sin\psi \hat{\psi} }{ ( r^4 \sin^2\varphi + \varsigma_+^2 \varsigma_-^2 \sin^2\psi )^{\frac{1}{2}}}.
\end{equation}
As we see, $ \hat\chi $ and $ \hat\varphi $ are not included in \eqref{invis}, which indicates that those directions and their corresponding angles remain in the reduced $ \mathbb{T}^3 $. 

On the other hand, from the $ \mathbb{R}^4 $ observer's point of view, the spin direction of $ \mathbb{T}^3 $ is directed toward $ \hat{\psi} $, and we expected that the direction $ \hat{\psi} $ to be invisible for the $ \mathbb{T}^3 $ observer. But, we see that the $ \mathbb{T}^3 $ observer can also detect the spin of the universe and somehow its direction. 

From $ \mathbb{R}^4 $ observer's point of view, the invisible direction for $ \mathbb{T}^3 $ can be obtained by solving the following relation,
\begin{eqnarray}
 \hat{n}_{\mathsf{vis}} = u_r \hat{r}+u_\varphi \hat{\phi} + u_\psi \hat{\psi} + u_\chi \hat{\chi}, \hspace{2cm}
 \hat{n}_{\mathsf{vis}} . \hat{n}_{\mathsf{invis}} = 0.
\end{eqnarray}
Therefore, the portion of the spin (according to the spherical coordinate $ \mathbb{R}^4 $) which is seen by the real observer is obtained as,
\begin{eqnarray}\label{nvis}
\hat{n}_{\mathsf{vis}} = \frac{\sqrt{1-u_{\varphi}^{2}-u_{\chi}^{2}}}{\sqrt{\varsigma_+^2 \varsigma_-^2 \sin^2 \psi + r^4 \sin^2 \varphi}}(\varsigma_+ \varsigma_- \sin\psi \hat{r} - r^2 \sin\varphi \hat{\psi}) +u_\varphi \hat{\varphi}+u_\chi \hat{\chi} , \quad  u_\varphi,u_\chi \in \mathbb{R}.
\end{eqnarray}
As we see, in the above direction, there is still a small portion for $ \hat{\psi} $, which means from the $ \mathbb{R}^4 $ observer's point of view, the $ \mathbb{T}^3 $ observer can see the spin vector $ \hat\Omega = \omega \hat{\psi} $.

 Nevertheless, the presence of free and arbitrary parameters $ u_\varphi , u_\chi  $ in \eqref{nvis}, indicates the fact that the embedded  $ \mathbb{T}^3 $ universe is isotropic according to its corresponding observer.

\subsubsection{Toward the spin direction and the fundamental constant $ \omega $}\label{Toward}
As we said before, for $ \mathbb{R}^4 $ observer, the spin of the embedded $ \mathbb{T}^3 $ is,
\begin{eqnarray}
\vec{\Omega}_{\mathsf{tot}} = \omega \hat{\psi},
\end{eqnarray}
and the observer in toroidal universe sees,
\begin{eqnarray}\label{omegavis}
&& \vec{\Omega}_{\mathsf{vis}} = \omega \hat{\psi} - \vec{\Omega}_{\mathsf{invis}} , \nonumber \\
&& \qquad =  \omega \hat{\psi} -(\omega \hat{\psi}.\hat{n}_{\mathsf{invis}})\hat{n}_{\mathsf{invis}} \nonumber \\
&& \qquad = \frac{r^2 \omega \sin\varphi}{r^4 \sin^2 \varphi + \varsigma^2_+ \varsigma^2_- \sin^2\psi}\left( r^2 \sin\varphi\hat{\psi}-\varsigma_+ \varsigma_- \sin \psi \hat{r}\right),
\end{eqnarray}
It should be noticed that $ \hat{r} $ and $ \hat{\psi} $ are $ \mathbb{R}^4 $ unit vectors. Thus, no one can claim that the spin direction and its value are completely observed by $ \mathbb{T}^3 $ observer.

Hence, to obtain the exact value, observed by the $ \mathbb{T}^3 $ observer, one should go on the constraints surface, i.e. $ \psi $ should be solved from $ \Phi_1 = 0 $. Then, by replacing the obtained  $ \psi $ in the \eqref{omegavis}, in the homogeneous limit of the universe, $ \epsilon \rightarrow 0 $, and for far or near distances (IR or UV), we would have,
\begin{eqnarray}
&& \vec{\Omega}_{\mathsf{vis}} . \hat{n}_{\mathsf{vis}} \biggr\vert_{\Phi_1}= \parallel \Omega_{\mathsf{vis}}^{(\mathbb{T}^3)} \parallel \nonumber \\
&&\hspace{2cm}  = \frac{r^2 \omega \sin\varphi \sqrt{1 - u_\varphi^2 - u_\chi^2}}{(r^4\sin^2\varphi + \varsigma^2 (1-\epsilon)^2 \sin^2 \psi)^{\frac{1}{2}}}\left( r^4 \sin^2\varphi + \varsigma^4(1-\epsilon)^2 \sin\psi \right) \biggr\vert_{\Phi_1} \nonumber \\
&& \hspace{2cm} = \frac{r^2 \omega \sin\varphi \sqrt{1 - u_\varphi^2 - u_\chi^2}}{(r^4\sin^2\varphi + \varsigma^2 (1-\epsilon)^2 \sin^2 \psi)^{\frac{1}{2}}}\biggr\vert_{\Phi_1}.
\end{eqnarray}

For the homogeneous limit of the universe, $ (\epsilon \rightarrow 0 ) $, and far distances, $ \varsigma=1, (\bar{r}-2)\rightarrow 0 $, the spin of the universe is obtained as,
\begin{eqnarray}
 \parallel \Omega_{\mathsf{vis}}^{(\mathbb{T}^3)} \parallel  \approx \omega (1-\frac{\epsilon}{8\sin^2\varphi})\sqrt{1 - u_\varphi^2 - u_\chi^2}.
\end{eqnarray}
Since the observer on $ \mathbb{T}^3 $, does not have any liberty to choose the $ \hat{\varphi} $ direction, one can have the average volume over the space as,
\begin{eqnarray}
dV^{(\mathbb{T}^3)} = r^2 \sin^2 \varphi \sin\psi dr d\varphi d\psi d\chi \biggr\vert_{\Phi_1} \sim \sin^2\varphi,
\end{eqnarray}
and vanish its dependence to the $ \varphi $,
\begin{eqnarray}\label{omega_star}
\parallel \Omega_{\mathsf{vis}}^{(\mathbb{T}^3)} \parallel^{\star}  \approx \omega (1-\alpha\epsilon)\sqrt{1 - u_\varphi^2 - u_\chi^2},
\end{eqnarray}
where, the star sign, $ \star $, indicates the spatial average in far distances, $ \left((\bar{r}-2)\rightarrow 0 \right) $, and $ \alpha $ is a numerical coefficient obtained via the spatial average volume integration,
\begin{eqnarray}
\alpha \sim \frac{1}{8} \int \frac{1}{\sin^2 \varphi} dV^{(\mathbb{T}^3)}
\end{eqnarray}
It is interesting to see the effect of inhomogeneity of the universe, $ \epsilon $,  on the fundamental constant of the universe's spin (or its Casimir), $ \omega $ in \eqref{omega_star}. In other words, $ \omega $ can be regarded as the bare spin in primordial field theories, without having the spacetime interaction or renormalization, and $ \parallel \Omega_{\mathsf{vis}}^{(\mathbb{T}^3)} \parallel^{\star} $ is the observable and renormalized spin. 

Similar to QED where mass and charge of the electron is corrected by the coefficient $ \alpha_{QED} = \frac{1}{137} $, in our model the intrinsic spin of the universe is also corrected via $ \epsilon $. This correspondence (or similarity) may intuit us to have a deeper understanding of the results of this model.

\begin{table}[h]
\resizebox{15cm}{!}{
\begin{tabular}{c c |c|c|c|c|c|c|}
  % after \\: \hline or \cline{col1-col2} \cline{col3-col4} ...
  \cline{3-8}
 & &  \multicolumn{6}{ c| }{Functions \& Objects} \\ \cline{3-8}
 & & Bare & Functionality & Observables & Resulted  & Fine Tuning  & Cut-off \\  
 & & Objects &  &  &  Model &  Number &  \\  \hline
 \multicolumn{1}{ |c  }{\multirow{6}{*}{\rotatebox[origin=c]{90}{Models}}}& \multicolumn{1}{ |c| }{QED} & electron  & renormalization & renormalized  & bare QED +  & $\alpha_{QED} = \frac{1}{137}$  & UV  \\  
\multicolumn{1}{ |c }{} & \multicolumn{1}{ |c| }{}  &  mass \& charge &  &  mass \& charge &  correction loops &   & energy \\  \cline{2-8}
\multicolumn{1}{ |c }{} &  \multicolumn{1}{ |c| }{Our}  &  &   &   & reduced  &  &  \\  
\multicolumn{1}{ |c }{} &   \multicolumn{1}{ |c| }{  Test}  & $ \vec{\omega} $ &  reduction   &  $ \parallel \Omega_{\mathsf{vis}}^{(\mathbb{T}^3)} \parallel^{\star} $ &       phase space  & $\epsilon$ & $\varsigma$ \\  
\multicolumn{1}{ |c }{} &   \multicolumn{1}{ |c| }{ Particle} &  & on $ \mathbb{T}^3 $  &   & of test particle &  &  \\  
\multicolumn{1}{ |c }{} &   \multicolumn{1}{ |c| }{} &  &   &   &  on $ \mathbb{T}^3 $ &  &  \\  \hline
\end{tabular}}
\caption{a correspondence (not an exact duality in a perfect sense) between a typical field theory and present particle model}
\end{table}

\subsection{Homogeneous torus universe}

In algebraic geometry presentation of the universe with topology of $ \mathbb{T}^3 $, from the $ \mathbb{R}^4 $ observer's point of view,  the radial component, $ r $, is finite.

Also, for the observer in $ \mathbb{R}^4 $, the UV region  $ 0 \leq r < \varsigma_+ - \varsigma_- $, especially the location $ r = 0 $ is not available by the $ \mathbb{T}^3 $ observer, and without imposing a fine tuning condition the overall universe will not be homogeneous. In other words, our visible universe with minimal length $ \epsilon $,
\begin{eqnarray}
\frac{\varsigma_+ - \varsigma_- }{\varsigma_+} = \epsilon \rightarrow 0,
\end{eqnarray}
can be considered as a homogeneous $ \mathbb{T}^3 $ universe, with infinitesimally unequal radii, and the parameter $ \epsilon $ controls the inhomogeneity.

Hence, we introduce a special scheme to introduce the 3 dimensional brane as a $ \mathbb{T}^3 $ shaped universe which is deviation from $ \mathbb{R}^4 $  with two almost equal radii, $ \varsigma_+ = \varsigma $ and $ \varsigma_- = (1 - \epsilon)\varsigma $. As we mentioned, the geometrical deformation and inhomogeneity of such universe is controlled by $ \epsilon $.

\section{Quantum cosmology investigations}
To study the infinitesimal structure of space by a test particle which is localized in a small region, quantum mechanics is a suitable tool. Although the test particle's coordinates which are used here is from the $ \mathbb{R}^4 $ observer's view, transforming to the particle's local coordinates  would be an easy task.

Therefore, the origin can be chosen by the local observer in $ \mathbb{T}^3 $, with the uncertainty of the order of  $ \varsigma \epsilon $. This observer can put his test particle in the origin and investigate the fine structure of the space, by studying its corresponding quantum mechanics. Also, he can find the proper estimate for the parameters $ \epsilon, \varsigma $ and $ \omega $ of the model \cite{Ida}.

Obtaining the symplectic algebra of the model, $ J_{\mu\nu}(z) $, the Heisenberg algebra of the main quantum observables in the UV limit can be obtained as,
\begin{eqnarray}
i \hslash J_{\mu\nu} \approx i \hslash(J_{\mu\nu}(0) + \epsilon J_{\mu\nu}(r) + O(\epsilon^2)),
\end{eqnarray}
where, $ J_{\mu\nu}(0) $ is the constant matrix with symplectic structure, and $J_{\mu\nu}(r) $ is the first order approximation of the symplectic structure for $ \epsilon = 0 $, which will be done only for the dimensionless radial component $ \bar{r} $.

Using Dirac brackets, the observer may make the quantum mechanics for the test particle to clearly describe the structure of the phase space of the model in quantum scale. Hence, by a good approximation, the Poisson structure $ J_{\mu\nu} $ can be quantized. 

By scaling the radius component of the model, $ r $, and its corresponding momenta, $ p_r $,  with the specific length $ \varsigma $ of the universe,
\begin{eqnarray}
\bar{r} = \frac{r}{\varsigma} ,	\qquad \bar{p}_r = \varsigma p_r,
\end{eqnarray}
one can have the quantized version of the model.

\begin{eqnarray}\label{JS}
&& \left[ \bar{r},\bar{p}_r\right]  = i \hslash \left( 1 - \bar{r}^2 (1+\epsilon) \sin^2 \varphi \right) ,\nonumber \\
&& [\bar{r},p_\psi]  =  \frac{- i \hslash}{2}  (2+\epsilon)  \bar{r}^2 \sin^2 \varphi , \nonumber \\
&& [\psi,\bar{p}_r] =- i \hslash \left(1 + \frac{1}{2} \epsilon + \frac{1}{8} \bar{r}^2 (1+\frac{3}{2}\epsilon) (1 - 8 \sin^2 \varphi)  \right), \nonumber \\
&& [\psi,p_\psi] = i \hslash (1 + \epsilon) \bar{r}^2 \sin^2\varphi , \nonumber \\
&& \left[ \bar{p}_r , p_\psi \right] = i \hbar \left( (1+\frac{1}{2}\epsilon) (\bar{r} \bar{p}_r+ \bar{p}_r \bar{r}) - 2 \bar{r} \varsigma^2 \omega \right) \sin^2 \varphi + O(\bar{r}^3 \bar{p}_r),  \nonumber \\ 
&& [\bar{p}_r,p_\varphi] = i\hslash \bar{r}^2  (1 + \frac{1}{2} \epsilon )\omega \sin 2\varphi , \nonumber \\
&& [p_\psi,p_\varphi]=  i\hslash \bar{r}^2 \varsigma^2 \omega \sin 2\varphi + O(\bar{r}^2 \bar{p}_r) .
\end{eqnarray}

Comparing commutators above with the common structure of quantum mechanics variables in pure 3-dimensional space of $ \mathbb{R}^3 $, one can see how the embedding procedure changes  the algebra of basic variables of the model. Also, we can see that the inhomogeneity of $ \mathbb{T}^3 $ is entered to the above algebra.

One can also compare these brackets with the non-rotating torus universe, previously modelled in \cite{torus}. We can see that except three last commutators, including $ \omega $, all commutators are  the same as before. Moreover, one can compare these results with similar type of Dirac brackets for particle on a torus knot in \cite{das}.

Moreover, although the test particle is confined on a 3-dimensional space, due to the embedding procedure, the above mentioned quantum commutators include $ \hat{\psi} $ and $ \hat{p}_\psi $ which can be regarded as hidden variables (like Einstein hidden variables) that  effect the quantum state of the particle.

On the other hand, as we see $ \omega $, which indicates the spin of $ \mathbb{T}^3 $, is appeared in three commutators. Two of them include $ \omega $ multiplied by the great radius of the universe, $ \varsigma $, which can be ignored by a renormalizing procedure which vanish infinities. In other words, we claim that those terms including $ \omega $ does not have any observable effect, but since they also include $ \hat{p}_\psi $, one can say that via those commutators, the momentum is injected form the hidden dimension of the universe to $ \mathbb{T}^3 $  and spins the brane.

Nevertheless, the $ \omega $ is appeared multiplied by $ \epsilon $ in the commutator of variables $ (\bar{p}_r,p_\varphi) $, and hence it can employ an observable  quantum mechanical effect. 

Also, all above mentioned commutators can be interpreted  in such a way where due to the embedding $ \mathbb{T}^3 $ and its spin in $ \mathbb{R}^4 $, the Planck constant is corrected. This claim can be investigated in high energy (tiny distances) regime. By the way, these forms of commutators in this model, inspire us to investigate IR/UV mixing \cite{Mann3}.

\subsection{Minimal length in $ \mathbb{T}^3 $ universe}

Like several works  \cite{Maggiore,Maggiore2,Maggiore3,Mann,Mann1,Mann2}, we see that in quantum commutators of our model \eqref{JS}, which are deviated from the common Heisenberg algebra, a minimal length is appeared. Hence, one could expand this symplectic algebra according to the dimension parameter of minimal length $ \epsilon $, by calculating them in quantum regime of small lengths, $ \bar{r} = \frac{r}{\varsigma} $ , and investigate such a claim\footnote{This means that such a fine tuning $ \epsilon \rightarrow 0 $ is existed in nature.}.

We think that the amount of inhomogeneity and deformation in this universe, $ \epsilon $ , can be estimated with the help of the cosmic data. Also, the fundamental length of this model can be regarded as the Hubble length, as $ R_H \approx 2 \varsigma N $, where $ N $ is the winding number on the periodic $ \mathbb{T}^3 $ \cite{nakahara2}.

The above-mentioned basic commutators help us to study the dynamics of the quantum test particle in a small localized region around the observer. This means that the test particle here acts as an elementary particle, like an electron, to test the effect of the topology and spin of the universe on the fine structure of the space.

On the other hand, as we will see further, this test particle can be regarded as a far universe (from the $ \mathbb{R}^4 $ observer that is coincided to the origin of $ \mathbb{T}^3 $ observer), where studying its dynamics may help us to understand the effect of the topology and spin of the universe on cosmic parameters.

The advantage of choosing the spherical coordinates to calculate basic operators in quantum Hilbert space is that one can easily obtain those commutators in the infinitesimal quantum distance approximations, by expanding the results around $ r = 0 $, where we can compare our results to the minimal lengths, and minimal momentum quantum mechanics\footnote{This approximation is claimed to be true, since at far distances the  particle's wave behaviour can be regarded negligible in comparison to its localization, and a quantum particle there can be regarded as a classical particle.} \cite{Mann,Mann1,Mann2}. Although models dealing with such phenomena, consider different issues such as correction and renormalization of Planck scale in high energy physics and their effects in quantum mechanics in curved spacetime, we consider the topology of the universe and the extrinsic geometry imposed via embedment, and also the spin of the universe to have the basic commutators changed. 

Thus, the quantum mechanics obtained from our model which is expressed by the basic commutators algebra \eqref{JS}, includes two fundamental constants, $ \varsigma $ and $ \omega $, and can be compared to \textit{Amelino-Commelia }models of \textit{doubly special relativity}, where it is based on the change in the particles' energy-momentum dispersion in small scales and high energies \cite{camelia,dsr,dsr1}. The symmetry properties and the algebra of symmetry generators of our model can be compared with \cite{Kawa1,Kawa2,Ruegg,Lukierski}. Moreover, our results can be compared to the models including extra fundamental parameters rather than $ \hslash $ and $ c $, such as those models studying the Lorentz symmetry below Planck scale \cite{simon,simon3,simon2}.

\subsection{Omitting the canonical couple $ (\hat{\psi},\hat{p}_\psi) $}

After quantizing a physical states in the formalism of constrained systems, solving the  Schr\"odinger equation solely can not completely determine all quantum states as $ H\mid phys> = E \mid phys> $. This happens due to the presence of gauge symmetry which is imported by first class constraints. Hence, one should have them in a more limited way of constrained equations as $ \hat{FC} \mid phys> =0 $, where $ \hat{FC} $ is the operator form of first class constraints \cite{torus}.

Hence, from the equivalence sets of states, first class constraints extract a complete set in the operator form. The operator form of constraints effects the states obtained from Hamiltonian, similar to the way of choosing a specific gauge and breaking the gauge symmetry.

In second class systems, we reduce the basic variables (and also quantum observable states), and with the help of those reduced variables and their algebra we find a proper representation for basic observables. Then we obtain the reduced Hamiltonian, $ H_{red} $, defined according to the reduced observables, and rewrite the Schr\"odinger equation as, $ H_{red} \mid phys> = E \mid phys> $, and then we can have the states.

The above mentioned method can be used only if by calculating Dirac brackets one can omit a canonical couple, such as two second class constraints as  $ [\hat{SC}_i,\hat{SC}_j] = i \hslash \delta_{ij} $.

For this model, one can omit the canonical couple $ (\hat{\psi},\hat{p}_\psi) $ by imposing the constraints, although their effect on the model would not vanish, e.g. the interaction of $ \hat{p}_\psi $ with other basic observables indicate that this operator conveys the intrinsic spin of the universe to the observable part of $ \mathbb{T}^3 $. Therefore, being embedded  in $ \mathbb{R}^4 $ does not only effects the $ \mathbb{T}^3 $ Hamiltonian, and it also effects the universe's observable representations. In other words, although this embedding is explained by second class constraints, it is also a gauge symmetry, and $ (\hat{\psi},\hat{p}_\psi) $ fixes this symmetry \cite{NC,hor,der}. \footnote{In the formalism of constrained systems, one can enter a second class couple as a first class constraint (gauge transformation generator), with a partner fixing the gauge symmetry \cite{heno}, such as here with $\psi$ as a first class constraint and $ p_\psi $ as its partner. Thus, although we have second class constraints, these constraints can be regarded as a fixed gauge symmetry (first class constraint). For instance, $ \Phi_1 $ is a first class constraint which is fixed consistently by $ \Phi_2 $. }

As we see in \eqref{JS}, $ \omega $ is not included in the commutator of $ (\hat{r},\hat{p}_r) $, and it is only appeared in the commutators of a phase space variable and one of angular momenta. Therefore one can conclude that although we got rid of operators ($ \hat{\psi},\hat{p}_\psi) $ by using constrained relations, their effect to the algebra is not negligible, since they have (specially $ \hat{p}_\psi $ has) a nonvanishing algebra with the basic observables of $ \mathbb{T}^3 $ . Hence, $ \omega $ can be regarded as a new Casimir which is entered to the algebra of observables of the model.

\section{Cosmic Dynamics}
Now, we can study the dynamics of a cosmic object, such as a galaxy, to detect the local and global structure of the space, by studying its Hamilton function and symplectic structure, and reading off the Hamilton equations of motion for far distance object as,
\begin{eqnarray}\label{hamilton}
 \dot{r} = \{r,H\}^* , \hspace{2cm}
 \dot{p_r} = \{p_r,H\}^*.
\end{eqnarray}

On a universe, where  the $ \mathbb{T}^3 $ observer does not sense any deviation or angular rotation, the test particle's energy is negligible in comparison to the parameters of the universe. As a mater of fact, the total energy of the universe is constructed via its geometrical structure, $ \varsigma $ , and movement, $ \omega $. Hence, we consider that the energy of a sample universe, such as our test particle, is almost zero,
\begin{equation}\label{E_cos}
E_{cosmos} \approx 0 \qquad, \qquad p_\varphi = p_\chi = 0,
\end{equation}

From the first Hamilton equations \eqref{hamilton}, one can obtain the relation between distance and velocity, (the Hubble law), and from the second one the acceleration of the universe is obtained.
\begin{eqnarray}
\dot{r} = v_r^{cosmos} = p_r (r,\varphi), \hspace{2cm}
\ddot{r} =  a_r^{cosmos}(r,p_r,\varphi).
\end{eqnarray}
Obtaining $ p_r $ from \eqref{E_cos}, we encounter several situations. In near distances, $ v_r^{cosmos} \sim r^2 $, and for distances near to the boundary of periodic universe $ \mathbb{T}^3 $, $ v_r^{cosmos} \sim r^{-\frac{3}{2}} $. Due to some observational evidences \footnote{We know from the Hubble law that we live in an epoch much further of inflation era (not to live in $ r \approx 0 $), and in the epoch before the deceleration phase (not to live in $ r \approx 2\varsigma $) \cite{Perlmutter}.}, extreme points, i.e. $ r \approx 0 $ and  $ r \approx 2\varsigma $ are put away. Hence, one can calculate around a median point, such as $ r \approx \varsigma $. Thus, for the velocity of the test particle with respect to the first order of $ \varsigma $, we have,
\begin{eqnarray}
&& v^{cosmos}=  \frac{2\omega \sin\phi }{3 \sqrt{3}  \left(4 \sin^2\phi +3\right)^{5/2}} (-16 \sin^4\phi (r (2 \text{$\epsilon $}+3)+4 \varsigma  (\text{$\epsilon $}-3))\nonumber \\
&& \hspace*{1.6cm} +24 \sin^2\phi  (r (3-13 \text{$\epsilon $})  +2 \varsigma  (5 \text{$\epsilon $}+3))+81 r ).
\end{eqnarray}
Also, for the acceleration to the zeroth order of $ \varsigma $, we obtain,
\begin{eqnarray}
a^{cosmos} = \frac{2\varsigma  \omega ^2 \sin ^2\phi}{(2 \cos 2\phi -5)^3} ((4-64 \text{$\epsilon $}) \cos 2\phi +64 \text{$\epsilon $}+2 \cos 4\phi -33).
\end{eqnarray}
The angle $ \varphi $ is a dynamical variable, but since we consider the universe in a particular time, we can think of a constant angle, having an average volume integration $ \bar{F} = \frac{2}{\pi}\int_0^{\pi} \sin^2\varphi F(\varphi) d\varphi $. Thus, we  have,
\begin{eqnarray}
&& v^{cosmos} = \varsigma \omega   (2.51  \text{$\epsilon $}+3.94) \times 10^{-1}+ \omega r (2.01 -4.48 \text{$\epsilon $})\times 10^{-1}, \label{va1}  \\
&& a^{cosmos} = \varsigma  \omega ^2 (1.16 -3.12 \text{$\epsilon $}) \times 10^{-1}\label{va2}.
\end{eqnarray}
The constant term in \eqref{va1}, which includes the $ \mathbb{T}^3 $ universe parameters $ \varsigma $, $ \omega $ and $ \epsilon $, is appeared due to the translation velocity of $ \mathbb{T}^3 $ universe, from the $ \mathbb{R}^4 $ observer. The second term, can be regarded as the Hubble law. As we see, the Hubble constant in this model is independent of the periodic radius of this universe, $ 2\varsigma $, and to this order of  approximation is only obtained via the spin of the universe (i.e. no spin, no Hubble law).

Here, the Hubble  constant is renormalized up to the first order of $ \epsilon $. This means that if we desire not to have a blue shift for the distances greater than the half of the radius of the universe, $ \epsilon < 0.45 $. Also, the positive acceleration limit, indicates a better limit of $ \epsilon < 0.37 $.

It is worth mentioning that the Hubble law gets a nonlinear form in the extreme point $ r = 2\varsigma $ and near boundaries as $ v_r^{cosmos} \sim (r-2\varsigma)^{\frac{-3}{2}} $, which is another testimony for our periodic $ \mathbb{T}^3 $ universe, embedded in $ \mathbb{R}^4 $.

In $ r \sim \varsigma $,  \eqref{va1} and  \eqref{va2} can be written in the form of its two characteristics of the model, i.e. $ \varsigma $ and $ \omega $ as \footnote{Here, $v_{\textit{ch}}$ and $ a_{\textit{ch}}$  indicate the characteristic velocity and acceleration respectively.},
\begin{eqnarray}
v_{ch}= \varsigma \omega \times10^{-1} \qquad, \qquad a_{ch} = \varsigma\omega^2  \times10^{-1} ,
\end{eqnarray}
and be corrected with the inhomogeneity parameter $ \epsilon $. 

Also,  velocity and acceleration of the universe in far distances $ (r \sim \varsigma) $, are obtained to the first order of inhomogeneity as,
\begin{eqnarray}
&& v(r) \approx v_{ch} ( \frac{r^2 }{\varsigma^2 }  (-2.52 \epsilon -4.48)+\frac{r}{\varsigma}   (0.56 \epsilon +10.98)+ (-0.01 \epsilon -0.54)) \\
&& a(r) \approx a_{ch} ( (6.25 -0.28 \epsilon)- \frac{r}{\varsigma}   (2.83 \epsilon +5.08)). \label{a_cos}
\end{eqnarray}
It is evident that in distances smaller than $ \varsigma $, the correction of the Hubble law gets smaller than the common form, and the acceleration decreases infinitesimally from a constant value.

Nevertheless, to study the expansion of the space for a distance $ \delta r \ll \varsigma $, (e.g. $ r = 0.01 \varsigma $) we have,
\begin{eqnarray}
\delta a  \approx - \frac{\delta r}{\varsigma}   (2.83 \text{$\epsilon $}+5.08)  a_{ch}.
\end{eqnarray}
Hence, the far universe decelerates, and this to much less than a percent of the constant acceleration we obtained before \eqref{va2}.

Therefore, the universe near the origin experiences the greater acceleration than a far one, and form the observer's point of view  the final state of the universe seems to have accelerating expansion.

Also, using the relation \eqref{a_cos}, for $ r = 0 $, we can obtain the classical big bang acceleration as,
\begin{equation}
a_{\textsf{bigbang}} = a_{ch}(6.25-0.28 \epsilon ).
\end{equation}

But, what about the acceleration near the boundary, $ (r = 2 \varsigma) $? As one can see, the universe experiences a big negative acceleration as,
\begin{eqnarray}\label{A_big}
a_{\textsf{big-crunch}} \approx - 5 a_{ch}(1+\epsilon)+ \frac{O(\epsilon^2)}{(r-2\varsigma)^2},
\end{eqnarray}
which happens due to the periodicity of $ \mathbb{T}^3 $ universe, and can be investigated in further works related to cyclic models \cite{cyclic,cyclic2,boun}.

As we see, \eqref{A_big} contains an infinite and singular term, which will be thrown away, and a negative one, which returns the far universe to the permitted area of $ r < 2 \varsigma $, where the second half period of a periodic universe appears. 

This return and having a negative acceleration in boundaries, can be regarded as a big crunch (in $ r \approx 2\varsigma $) that happens in a periodic universe after a big bang (in $ r = 0 $).

Thus, with the help of embedding approach and constrained formalism, without using field theory or general relativity and its corrections such as $ f(R) $ gravity, we obtained a big crunch phase for our model, and all that is done only by embedding a $ \mathbb{T}^3 $ in $ \mathbb{R}^4 $, without considering a potential or a cosmological constant 
\cite{f_r}.

%On the other hand, due to homogeneity of the space $ \mathbb{T}^3 $, the point  $ r \approx 2\varsigma $ is a singular point, since $ r \approx 2\varsigma $ is a physical distance between the origin and boundary of the universe.

%But, if someone claims that the singularity $ r = 0 $ is a coordinatial singularity and can be vanished by the translation of the origin, here we encountered two singularities in acceleration in $ r \approx 0 $ and  $ r \approx 2\varsigma $, where at least one of them cannot be eliminated. Hence, if the observer in $ r \approx 0 $ observes the singularity and bouncing in $ r \approx 2\varsigma $, by translating him to $ r' \approx 2\varsigma $, he observes the singularity and bouncing in $ r' \approx 0 $.

\subsection{Some possible phenomenological and numerical investigations}
Our cosmological model here is based upon the particle nature (not the field) and its interactions (in quantum scale of a spinless, chargeless and without any other quantum feature of quantum particles, containing only mass, unit inertia, and in large scales as a far universe with the negligible energy in comparison to the energy of the universe) in a background space (not spacetime).

Since we did not use any general relativistic and field theory formalism, no cosmological constant is entered to our model. Anyway, our test particle (which is an observing far universe) experiences an acceleration for the universe. As we showed, we obtained a linear expansion of a positive acceleration over the median point of a periodic universe. Fitting this acceleration with observational data, one can tune $ a_{ch} $ and find its proper limit.

Moreover, since we are in a positive acceleration era, the positivity condition imposes a limit on the inhomogeneity parameter of the universe. Also, we have another limit which is imposed on $ \epsilon $ by the Hubble law of our model. Fitting this parameter with observational data imposes a limit on $ v_{ch} $, according to the half radius of the periodic universe. One should notice that none of those fittings will provide an absolute value for $ \varsigma $.

But, if we assume that there exist a fine tuning between two radii of the $ \mathbb{T}^3 $, the quantum structure of spacetime introduces a fundamental minimal length as $ \varsigma_+ - \varsigma_- = \epsilon\varsigma $. Hence, one can estimate the order of magnitude of $ \varsigma $ by studying the algebra of quantum commutators in \eqref{JS}, which will be important in at the order of Planck length. 

Thus, by extracting the minimal length of noncommutative algebra of quantum commutators in our model, and comparing them with the quantum gravity specific length (Planck length), one can find a proper limit of $ \varsigma $. In addition, the above mentioned algebra which includes $ \epsilon \omega $ also provides a minimal momentum, that may help us to impose limits for $ \varsigma $, $ \epsilon $ and  $ \omega $.

%On the other hand, one should tune at least one of the parameters $ \varsigma $, $ \epsilon $ and  $ \omega $,  by extracting the proper quantum gravity (i.e. finding proper presentations for algebra and writing the Schr\"odinger equation), via free particle's quantum mechanical investigation such as Hydrogen atom, or harmonic oscillator (as a basis for every field theoretical expansions), without using the observational data and quantum gravity limits. Here, we claim that estimating $ \epsilon $ by the sign of acceleration and velocity is not a suitable approach, since the quantum gravity length is as the order of $ l_p \sim 10^{-35} m $ and the universe's radius is as the the order of $ D \sim 10^{26} $, and the fine tuning in our model estimates $ \epsilon \sim 10^{-61} $.

Also, studying the quantum harmonic oscillator of the model may help us to relate the model to the CMB fluctuations.

\section{Discussion}

In this article we modelled a spinning, finite and periodic cosmological universe based on a $ \mathbb{T}^3 $, embedded in a $ \mathbb{R}^4 $ space, where its inhomogeneity and deformation is controlled via the dimensionless parameter $ \epsilon $, as the difference of two torus' radii. 

The spin vector is chosen in a way not to enter a  fundamental or  coordinate singularity to the Hamiltonian function, and at the same time adds the fundamental constant angular velocity, $ \omega $, to the model. Also, we showed that that the presence of $ \omega $ produces a scalar and a magnetic vector potential, which  can be interpreted as the magnetic background resulting the constant noncommutativity. This magnetic background changes the symplectic structure of the phase space of this model to the  the symplectic structure of a noncommutative one, where a part of this noncommutativity is controlled via $ \omega $, and the other part, as it has been shown in \cite{torus}, is due to the Casimirs, i.e. $ \varsigma_{1} $ and $ \varsigma_{2} $ , radii of  $ \mathbb{T}^3 $. 

One should notice that we assumed two radii of our toroidal universe to be nearly equal, then our model satisfies the homogeneity condition by approximation. We also showed how the homogeneity of the $ \mathbb{T}^3 $ universe, with infinitesimally unequal radii, is controlled by the parameter $ \epsilon $.  Also,  we found the non-unique spin direction as a preferable direction in space, including some free parameters, which control the universe's anisotropy. Hence, with the help of that and assuming the isotropic condition of the universe, we obtained the invisible spin direction for $ \mathbb{T}^3 $, from the $ \mathbb{R}^4 $ observer's point of view.

Then, by going to the constrained surface and reducing a couple of phase space coordinates, we imported the constraints into the dynamics of the system consistently.

Afterwards, by using Dirac brackets and quantizing Poisson structure of the model, we clearly describe the structure of the phase space of the model in quantum scale. This is the starting point of making the quantum mechanics as \eqref{JS}, which is the noncommutative fundamental quantum commutators of the model, although we did not present any representation for these fundamental operators. Quantum commutators obtained via this procedure show that the embedding procedure changes the algebra of basic variables of the model and the inhomogeneity of $ \mathbb{T}^3 $ is entered to the obtained algebra.

The form of commutators indicate that $ \omega $ can employ an observable quantum mechanical effects and  in our model the momentum is injected form the hidden dimension of the universe to the $ \mathbb{T}^3 $, which spins the brane. This can be interpreted  in such a way that due to the embedding $ \mathbb{T}^3 $ and its spin in $ \mathbb{R}^4 $, the Planck constant is corrected. This claim can be investigated in high energy regime and inspires us to investigate IR/UV mixing. We can also see that  quantum commutators include a minimal length and minimal momentum, and thus the model can be compared to some models based on doubly special relativity.

Calculating the velocity of the test particle, we obtained the corresponding Hubble law in our model. As we see, the Hubble constant which is normalized up to the first order of $ \epsilon $, is independent of the periodic radius of this universe, $ 2\varsigma $, and only depends to the spin of the universe.

Moreover, to study the expansion of the space in different distances we calculated the acceleration and its variation, which shows that the universe near the origin experiences the greater acceleration than a far one, and form the observer's point of view  the final state of the universe seems to experience an accelerating expansion. This would be an important result since we obtained the acceleration of the universe without considering a cosmological constant. 

Also,  we calculated a positive classical acceleration at $ ( r = 0) $ and a big deceleration near the boundary, $ (r = 2 \varsigma) $. Hence, one can say that in origin of the model we have a big bang and after a half period in $ r \approx 2\varsigma $ we encounter a big crunch which is the feature of a periodic universe. We hope that this feature of the model can be investigated in further works related to cyclic  and bouncing cosmologies.

%\appendix
%\section{Some title}
%Please always give a title also for appendices.

\acknowledgments
The authors acknowledges support from CAPES under the program PNPD during the completion of this work.

%\paragraph{Note added.} This is also a good position for notes added after the paper has been written.

% BIBLIOGRAPHY
% use BIBTEX if you want
%\bibliographystyle{JHEP}
%\bibliography{yourBIBfiles}

% The bibliography will probably be heavily edited during typesetting.
% We'll parse it and, using the arxiv number or the journal data, will
% query inspire, trying to verify the data (this will probalby spot
% eventual typos) and retrive the document DOI and eventual errata.
% We however suggest to always provide author, title and journal data:
% in short all the informations that clearly identify a document.

\end{document}